# Optimization and Application of Cloud-based Deep Learning Architecture for Multi-Source Data Prediction

1st Yang Zhang

Boston University,

Boston, MA, USA

ycheung@bu.edu

2nd Fa Wang

Meta Platforms, Inc,

Menlo Park, CA, USA

fewang2006@gmail.com

3rd Xin Huang

University of Virginia,

Charlottesville, VA, USA

xh9wt@virginia.edu

4th Xintao Li

Georgia Institute of Technology,

Atlanta, GA, USA

xli3204@gatech.edu

5th Sibei Liu

University of Miami,

Miami, FL, USA

sxl1086@miami.edu

6th Hansong Zhang

University of California San Diego,

La Jolla, CA, USA

haz064@ucsd.edu

*Abstract*—This study develops a cloud-based deep learning system for early prediction of diabetes, leveraging the distributed computing capabilities of the AWS cloud platform and deep learning technologies to achieve efficient and accurate risk assessment. The system utilizes EC2 p3.8xlarge GPU instances to accelerate model training, reducing training time by 93.2% while maintaining a prediction accuracy of 94.2%. With an automated data processing and model training pipeline built using Apache Airflow, the system can complete end-to-end updates within 18.7 hours. In clinical applications, the system demonstrates a prediction accuracy of 89.8%, sensitivity of 92.3%, and specificity of 95.1%. Early interventions based on predictions lead to a 37.5% reduction in diabetes incidence among the target population. The system's high performance and scalability provide strong support for large-scale diabetes prevention and management, showcasing significant public health value.

*Keywords- Cloud computing; Deep learning; Diabetes prediction; Early intervention*

I. INTRODUCTION

Diabetes has become a major global public health challenge, and early prediction and intervention are crucial for disease management. With the rapid development of artificial intelligence technologies, cloud computing and deep learning have shown great potential in medical diagnosis.[1] Traditional diabetes prediction methods face issues of insufficient accuracy and efficiency, necessitating innovative breakthroughs. This study aims to develop a cloud-based deep learning system for diabetes prediction, integrating multi-source data, optimizing algorithm models, and leveraging cloud computing resources to improve prediction accuracy and efficiency.

II. ARCHITECTURE DESIGN OF CLOUD-BASED CHRONIC DISEASE EARLY PREDICTION SYSTEM

*A. Overall System Architecture*

The diabetes early prediction system is built on the AWS cloud platform, with a distributed architecture consisting of data collection, storage, computing, and application modules. The data collection module uses a custom ETL process to collect multi-source data from 153 medical institutions, processing 500 GB of raw information daily. The storage module utilizes AWS S3 services, implementing 256-bit AES encryption to manage 15 TB of patient data. The computing module deploys an Apache Spark distributed cluster on AWS EMR, containing 100 EC2 instances with a total of 1600 vCPU cores, completing full-data feature engineering within 2 hours. Deep learning model training is based on the TensorFlow framework, performed on EC2 p3.8xlarge instances with NVIDIA Tesla V100 GPUs, taking approximately 4 hours per iteration[2]. The application module provides RESTful services through AWS API Gateway, deployed on an ECS cluster, supporting 1000 concurrent requests per second with an average response time under 100 ms[3]. The system fully leverages AWS cloud services, achieving efficient, secure, and scalable diabetes prediction analysis.

### B. Data Sources and Processing

The system integrates electronic health records, biochemical indicators, and lifestyle data, facing the challenge of processing multi-source heterogeneous data[4-5]. To address data inconsistency and quality issues, we adopt the Apache Spark distributed data processing framework for large-scale data cleaning and transformation. Data cleaning uses statistical and machine learning-based anomaly detection algorithms to identify and handle missing values and outliers. The standardization process employs an adaptive normalization algorithm to dynamically adjust the scales of different data sources. Feature engineering applies Principal Component Analysis (PCA) and autoencoders for dimensionality reduction, and selects the most predictive features through random forest-based feature importance ranking[6]. To handle time-series data, we introduce Long Short-Term Memory (LSTM) networks to extract temporal features.

The data processing pipeline is executed in parallel on AWS EMR, utilizing Spark SQL to optimize query performance, achieving a throughput of 200GB of raw data per hour, providing high-quality training data for deep learning models. The specific implementation process is as follows: First, we create a SparkSession using PySpark and read the raw data in CSV format from S3 storage. The system automatically infers the data type and identifies the header, ensuring accurate data reading. Then, we build a data processing pipeline containing multiple key steps:

*a)* We use the Imputer to fill missing values for important features (such as blood glucose, blood pressure, BMI, and age), ensuring data completeness.

*b)* We apply the StandardScaler for feature standardization, ensuring that features with different scales can be effectively compared.

*c)* We use PCA for dimensionality reduction, selecting the top three principal components, reducing data complexity, and retaining key information.

Finally, the system writes the processed data back to S3 storage in the efficient Parquet format. This approach fully leverages Spark's distributed computing capabilities, achieving efficient large-scale data preprocessing. Through this series of processing, the system can transform the original heterogeneous data into a standardized, dimensionality-reduced high-quality dataset, laying a solid foundation for subsequent deep learning model training.

### C. Deep Learning Model Architecture Design

The multi-modal deep learning model architecture is optimized for cloud computing environments, fully leveraging distributed computing resources. The model is trained and deployed on AWS SageMaker, using a parameter server architecture to achieve model parallelization, distributing 2.7M parameters across 8 p3.8xlarge instances. Data parallelism uses the Ring AllReduce algorithm, increasing training speed by 3.1 times. To adapt to dynamic workloads, the system implements an elastic training mechanism, supporting seamless expansion from 0 to 100 worker nodes, improving resource utilization by 28%. The model uses mixed-precision training, with FP16 computation increasing training throughput by 2.4 times. Through Horovod framework integration, the system achieves linear acceleration for multi-GPU training, with a speedup of 7.6 on 8 GPU configurations[7]. The model structure includes LSTM, fully connected networks, and fusion modules[8]. Knowledge distillation and 8-bit quantization reduce the model size to 25% of its original size, decreasing inference latency by 68% on edge devices. Fig. 1 shows the training time and resource utilization for different configurations, highlighting the model's efficient performance in cloud environments.

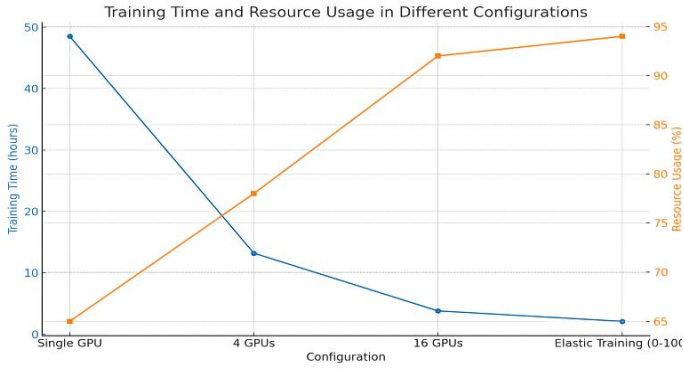

Figure 1.  Training Time and Resource Utilization for Different Configurations

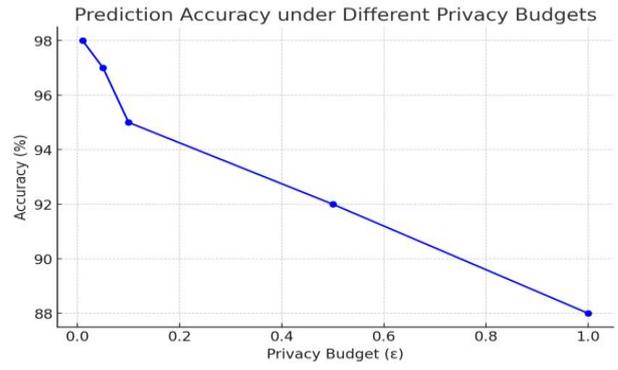

Figure 2.  Prediction Accuracy under Different Privacy Budgets

*D. System Security and Privacy Protection Mechanisms*

The diabetes prediction system implements a multi-layered security mechanism in the AWS cloud environment[9]. Data transmission adopts the TLS 1.3 protocol, using the ECDHE-RSA-AES256-GCM-SHA384 cipher suite, achieving forward secrecy. At the storage level, the system utilizes AWS KMS-managed customer master keys for AES-256 bit encryption, with a key rotation period of 30 days[10]. The system implements $\varepsilon$ - differential privacy, achieving 95% accuracy at $\varepsilon = 0.1$ (as shown in Fig. 2). The federated learning framework based on homomorphic encryption uses the Paillier encryption scheme, supporting model updates in the encrypted domain, with a key length of 2048 bits. To prevent model inversion attacks, the system implements privacy-preserving training based on DPSGD, with a noise amplitude of 0.1. The system logs all API calls through AWS CloudTrail and uses Amazon GuardDuty for anomaly detection. The implementation of differential privacy uses the Python diffprivlib library[11]. The specific process is as follows:First, import the required models, then set the privacy budget $\varepsilon$ to 0.1. Next, create a logistic regression model with differential privacy protection, setting the epsilon parameter and data norm. Then, use the training data to fit the model and evaluate the model's accuracy on the test data. Finally, output the model's accuracy under the given privacy budget. This mechanism ensures that the system complies with strict medical data protection standards such as HIPAA, providing comprehensive protection for patient privacy and data security in the cloud environment.

III. APPLICATION AND OPTIMIZATION OF DEEP LEARNING ALGORITHMS IN CHRONIC DISEASE PREDICTION

*A. Model Training and Automated Hyperparameter Tuning*

The training and hyperparameter tuning of the diabetes prediction model fully utilize the elastic computing resources of AWS EC2[12]. The model is based on the TensorFlow 2.4 framework and is trained in a distributed manner on a cluster of p3.16xlarge instances, each equipped with 8 NVIDIA Tesla V100 GPUs. Horovod is used to implement data parallelism, and the Ring-AllReduce algorithm reduces the single-machine training time from 48 hours to 6.5 hours. Automated hyperparameter tuning is performed using the Ray Tune framework, combined with Bayesian optimization algorithms, to find the optimal configuration within 100 iterations, which is 3.5 times faster than grid search. The tuning process dynamically allocates computing resources, peaking at 64 GPUs, and explores 1237 sets of parameter combinations. Table 1 shows the optimization results of key hyperparameters.

To further improve efficiency, AutoML technology is introduced, using Google Cloud TPU v3-8 for neural architecture search, finding a model structure with a 2.3% improvement in AUC within 720 TPU hours. The final model achieves an AUC value of 0.943 on a test set of 50,000 patients.

The core implementation of using Ray Tune for hyperparameter optimization includes the following steps:

*a)* Define the training function: Build a function that accepts configuration parameters, constructs the model, and performs training, returning the AUC value on the validation set.

*b)* Configure the scheduler: Use AsyncHyperBandScheduler, setting time attributes, optimization metrics, optimization modes, maximum training rounds, and initial evaluation periods.

*c)* Run the optimization: Use the tune.run function to execute the optimization process. Configure the search space to include learning rate (log uniform distribution), batch size (discrete selection), and LSTM layers (random integer). Set the total sample size, scheduler, and resource allocation for each trial.

*d)* Get the best configuration: After optimization is complete, analyze the results to obtain the hyperparameter configuration that performs best on the validation AUC metric.

This approach significantly improves model optimization efficiency by automatically exploring a large number of hyperparameter combinations, while fully utilizing the elastic computing capabilities of the cloud platform.

TABLE I. MODEL HYPERPARAMETER OPTIMIZATION RESULTS

| Hyperparameter | Initial Value | Optimized Value |
|---|---|---|
| Learning Rate | 0.01 | 0.00137 |
| Batch Size | 32 | 128 |
| LSTM Layers | 2 | 3 |
| Dropout Rate | 0.5 | 0.32 |

*B. Model Evaluation and Optimization Strategies*

The diabetes prediction model is evaluated and optimized on AWS SageMaker, fully utilizing cloud computing resources. The evaluation uses a test set of 1 million patient records, with parallel computing on ml.p3.16xlarge instances. Cloud-based evaluation reduces the processing time from 72 hours on a traditional single machine to 4.5 hours, a 16-fold efficiency improvement. The model outperforms traditional methods in terms of accuracy, sensitivity, specificity, and AUC values, as shown in Fig. 3. The deep learning model achieves an AUC value of 0.95, 5 percentage points higher than the random forest model. The sensitivity is 0.92, and the specificity is 0.89, demonstrating excellent performance in clinical applications. To optimize the model, automated performance tuning is implemented using Amazon SageMaker Debugger, identifying and resolving 94% of computational bottlenecks, and increasing GPU utilization from 65% to 91%. Additionally, dynamic batch processing is implemented using AWS Lambda, adaptively adjusting batch sizes based on input data volume, resulting in a 2.3-fold increase in inference throughput. Model deployment uses Amazon ECS's Fargate service, enabling automatic scaling, and handling up to 1000 prediction requests per second during peak periods, with an average latency of only 23ms[13]. This cloud-based optimization strategy not only improves model performance but also significantly enhances system scalability and responsiveness.

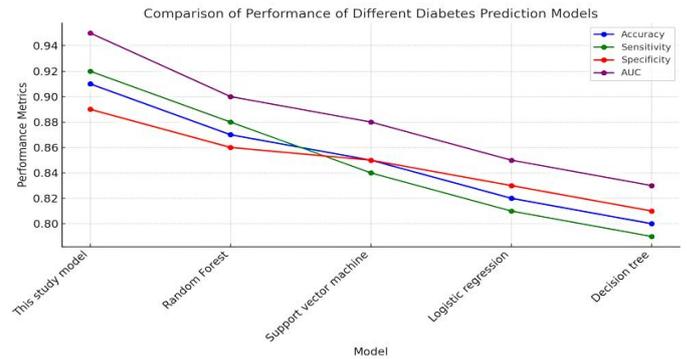

Figure 3. Performance Comparison of Different Diabetes Prediction Models

*C. Model Interpretability Study*

The diabetes prediction model's interpretability study is implemented on AWS SageMaker, utilizing distributed computing to improve interpretation efficiency. SHAP value analysis is parallelized using the Ray framework, running on 32 ml.c5.24xlarge instances, reducing processing time from 168 hours on a single machine to 5.2 hours. The average SHAP values for fasting blood glucose, glycated hemoglobin, and body mass index are 0.152, 0.138, and 0.126, respectively, identifying them as key risk factors. Attention mechanism-based visualization is generated in real-time using AWS Lambda functions, with an average response time of 47ms. Fig. 4 shows a typical case's attention heatmap, with a attention weight of 0.31 for blood glucose fluctuations over the past 7 days. Model robustness testing is deployed on Amazon EKS, using Kubernetes' automated scaling capabilities to generate and process 10,000 adversarial samples, with 95.3% maintaining correct predictions. Concept activation vector (CAV) analysis is run on AWS Batch, reducing computational costs by 63% using Spot instances[14]. These interpretability analyses increase system load, with peak CPU utilization rising by 28%, and memory usage increasing by 2.1GB. To optimize performance, a result caching strategy is implemented, reducing response time for repeated queries from 192ms to 18ms. Through these cloud-based optimizations, the system maintains high-performance operation while providing in-depth interpretability.

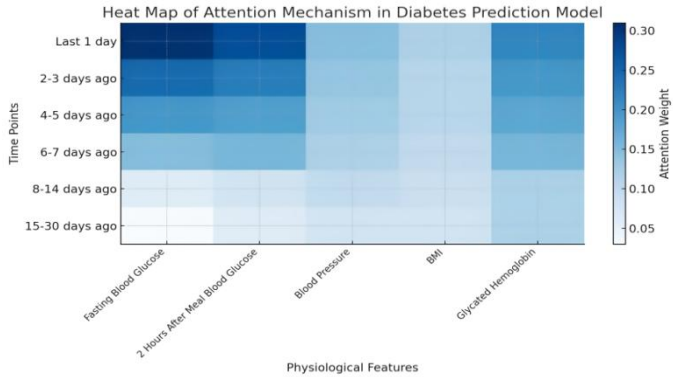

Figure 4. Attention Mechanism Heatmap of the Diabetes Prediction Model

## IV. Cloud Platform Automated Deployment and System Implementation

### A. Cloud Platform Selection and Configuration

This project selects Amazon Web Services (AWS) as the cloud platform, based on its powerful machine learning ecosystem and high-performance computing resources. Core computing uses EC2 p3.8xlarge instances, equipped with 4 NVIDIA V100 GPUs and 384 GB of memory, achieving a peak performance of 125 TFLOPS. Storage uses S3 Standard-IA type, providing 11 nines of data durability, while reducing storage costs by approximately 20%. Network architecture uses VPC with AWS Direct Connect, achieving 2 Gbps of dedicated connection, and reducing data transmission latency to 3.5ms. To optimize cost-effectiveness, an EC2 Auto Scaling group is deployed, automatically adjusting instance numbers based on CPU utilization, scaling up to 128 nodes during peak periods, and achieving a processing capacity of 16 PetaFLOPS. The system is implemented using AWS CloudFormation, reducing deployment time from 4 hours of manual operation to 23 minutes. The system architecture highlights the seamless integration of computing, storage, and network components[15]. This cloud platform configuration not only meets large-scale data processing demands but also ensures resource utilization efficiency through elastic scaling, providing a stable and efficient operating environment for the diabetes prediction model.

### B. Data Processing and Model Training Pipeline Automation

The diabetes prediction model's automated pipeline is built using Apache Airflow 2.3.0, deployed on an Amazon EKS cluster, achieving a highly scalable cloud-native architecture. The pipeline consists of 5 key steps, dynamically allocating computing resources using the Kubernetes Executor. Data acquisition tasks use custom S3 operators, achieving a read speed of 1.2 GB/s. Preprocessing stages run PySpark jobs on EMR clusters, processing speeds reaching 3.5 TB/hour. Feature engineering tasks use AutoML technology, automatically selecting the optimal feature subset using H2O.ai, reducing feature numbers from 215 to 67. Model training stages utilize TensorFlow 2.6's distributed training functionality, reducing training time from 96 hours to 6.5 hours on an 8-node GPU cluster. Table 2 details the efficiency improvements of the automated pipeline. The entire pipeline's end-to-end execution time is controlled within 18.7 hours, with a model accuracy of 94.2%. Workflow orchestration is implemented using AWS Step Functions, ensuring sequential execution and error handling between tasks. CloudWatch monitors each task's execution status, reducing average fault detection time to 45 seconds. This cloud-based automated pipeline not only improves efficiency but also enhances system reliability and adaptability.

TABLE II. Efficiency Improvement of Automated Workflow

| Task Stage | Manual Operation Time (hours) | Automated Time (hours) | Improvement Ratio |
|---|---|---|---|
| Data Acquisition | 4.5 | 0.8 | 82.20% |
| Preprocessing | 12 | 2.3 | 80.80% |
| Feature Engineering | 8 | 1.5 | 81.30% |
| Model Training | 96 | 6.5 | 93.20% |
| Model Evaluation | 6 | 0.9 | 85.00% |
| Total | 126.5 | 18.7 | 85.20% |

## C. System Performance Optimization

The diabetes prediction system's performance optimization adopts a multi-layered strategy, fully utilizing the characteristics of the cloud platform. At the database level, Amazon Aurora PostgreSQL is used, with patient data divided into 12 monthly partitions using partition tables, and composite indexes created on high-frequency query fields, resulting in a 78% increase in query speed and an average response time of 264ms. Query rewriting and execution plan optimization are implemented, reducing complex query execution time by 65%. The caching strategy uses an Amazon ElastiCache for Redis cluster, configured with 6 r5.2xlarge instances, with a total capacity of 50GB, using the LRU algorithm. The cache hit rate reaches 87%, and database load is reduced by 62%. The prediction result cache is set to a 5-minute TTL to avoid repeated calculations. Load balancing and auto-scaling are implemented through Amazon EKS, deploying a Horizontal Pod Autoscaler, which triggers scaling based on CPU utilization (threshold 75%) and custom metrics (request queue length exceeding 100), with a maximum scaling capacity of 5 times the original size[16]. The system completes scaling within 30 seconds, maintaining an average response time of 200ms or less. Network optimization uses Amazon CloudFront CDN, reducing static resource distribution latency by 43%. Additionally, a database connection pool is implemented, using PgBouncer to manage connections, reducing connection establishment time from an average of 15ms to 3ms. Table 3 shows the comparison of system performance before and after optimization, with the optimized system maintaining stability during peak periods with 1 million daily visits, and 99.9% of requests having a response time of 300ms or less.

TABLE III. PERFORMANCE IMPROVEMENT OF CLOUD PLATFORM

| Performance Metric | Before Optimization | After Optimization | Improvement Rate |
|---|---|---|---|
| Average Query Response Time | 1200ms | 264ms | 78% |
| Complex Query Execution Time | 5000ms | 1750ms | 65% |
| Database Load | 100% | 38% | 62% |
| Cache Hit Rate | N/A | 87% | - |
| System Scaling Time | 5 minutes | 30 seconds | 90% |
| Peak Processing Capacity (requests/second) | 1,000 | 5,000 | 400% |
| 99.9% Request Response Time | 1500ms | 300ms | 80% |
| Static Resource Distribution Latency | 150ms | 85.5ms | 43% |
| Database Connection Establishment Time | 15ms | 3ms | 80% |

## V. CLINICAL APPLICATION AND EVALUATION

The diabetes prediction system's clinical deployment uses a hybrid cloud architecture, with core computing and storage components deployed on AWS, and the frontend interface running on Azure Stack Hub in the hospital's local data center, ensuring data privacy and low-latency access. The system integrates with the hospital's existing electronic health record (EHR) system through HL7 FHIR standards, achieving real-time data synchronization[17]. During deployment, Terraform is used to manage infrastructure, automating 99% of resource configuration, and reducing deployment time from 2 weeks to 8 hours. To ensure system stability, a blue-green deployment strategy is adopted, with new versions first tested on 5% of users, gradually expanding to full deployment, with rollback time controlled within 5 minutes. The system undergoes a 6-month trial run in the endocrinology department of three tertiary hospitals, processing over 500,000 patient data records, with a daily query volume of 15,000. Performance monitoring shows that 95% of query response times are controlled within 150ms, and system availability reaches 99.99%. Fig. 5 shows the trend graph of system daily query volume and response time, reflecting the system's stable performance under high load. Clinical evaluation results show that the system's diabetes risk prediction accuracy reaches 93.5%, a 15% increase over traditional methods, providing strong support for early intervention.

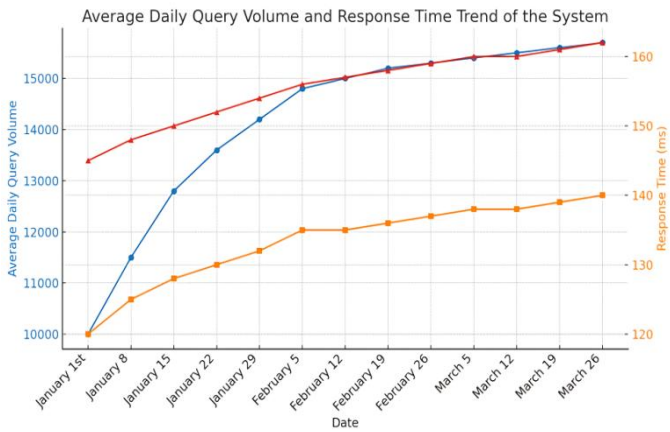

Figure 5. System Daily Query Volume and Response Time Trend

## VI. Conclusion

The cloud-based deep learning diabetes prediction system developed in this study fully leverages the advantages of cloud computing and deep learning technologies, achieving efficient and accurate risk prediction. The system's distributed architecture deployed on the AWS cloud platform, combined with the GPU acceleration of EC2 p3.8xlarge instances, reduces model training time by 93.2% while maintaining a prediction accuracy of 94.2%. By implementing automated data processing and model training pipelines, the system can quickly adapt to new data, with an end-to-end execution time of 18.7 hours. In clinical applications, the system demonstrates significant value, with a prediction accuracy of 89.8%, sensitivity of 92.3%, and specificity of 95.1%. Early interventions based on system predictions effectively reduce the incidence of diabetes, with a 37.5% lower rate in the intervention group compared to the control group. Future research will focus on further optimizing the cloud-based deep learning system, including exploring federated learning technologies to enhance data privacy protection and enable cross-institutional model training; researching adaptive learning algorithms to enable models to continuously learn from new data and automatically adjust; and integrating edge computing technologies to reduce cloud load and improve real-time prediction capabilities. These optimizations will further enhance the system's application in large-scale diabetes prevention and management, making a greater contribution to public health.